\documentclass[aps, twocolumn]{revtex4}
\usepackage{amssymb}
\usepackage{amsmath}
\usepackage{epsfig}
\begin{document}

%




\title{Is Fault-Tolerant Quantum Computation Really Possible? }                           

\author{M. I. Dyakonov}   

\affiliation{Laboratoire de Physique Th\'eorique et Astroparticules,
Universit\'e Montpellier II, France}

\vskip 2cm



\begin{abstract}

The so-called "threshold" theorem says that, once the error rate per qubit per gate 
is below a certain value, indefinitely long quantum computation becomes feasible, 
even if all of the qubits involved are subject to relaxation processes, and all the 
manipulations with qubits are not exact. The purpose of this article, intended for 
physicists, is to outline the ideas of quantum error correction and to take a look 
at the proposed technical instruction for fault-tolerant quantum computation. It 
seems that the mathematics behind the threshold theorem is somewhat detached from 
the physical reality, and that some ideal elements are always present in the 
construction. This raises serious doubts about the possibility of large scale 
quantum computations, even as a matter of principle.

\end{abstract}

\maketitle


{\bf 1. Introduction}\\

The answer that the quantum computing community currently gives to this question is
a cheerful "yes".  The so-called "threshold" theorem says that, once the error rate 
per qubit per gate is below a certain value, estimated as $10^{-4} - 10^{-6}$, 
indefinitely long quantum computation becomes feasible, even if all of the $10^{3}-10^
{6}$ qubits involved are subject to relaxation processes, and all the manipulations 
with qubits are not exact.  By active intervention, errors caused by decoherence can 
be detected and corrected during the computation.  Though today we may be several 
orders of magnitude above the required threshold, quantum engineers may achieve it 
tomorrow (or in a thousand years). Anyway large-scale quantum computation is possible in 
{\it principle}, and we should work hard to achieve this goal.

The enormous literature devoted to this subject (Google gives 29300 hits for 
"fault-tolerant quantum computation") is purely mathematical. It is mostly produced by 
computer scientists with a limited understanding of physics and a somewhat restricted 
perception of quantum mechanics as nothing more than unitary transformations in 
Hilbert space plus "entanglement".  On the other hand, the heavy machinery of the 
theoretical quantum computation with its specific terminology, lemmas, {\it etc}, is not 
readily accessible to most physicists, including myself. The vast majority of 
researchers, who start their articles with the standard mantra that their (whatever) 
subject is pertinent to quantum computation, do not really understand, nor care, 
what stands behind the threshold theorem and how quantum error correction is supposed 
to work. They simply accept these things as a proven justification of their activity.

Meanwhile, even the most ardent proponents of quantum computing recognize today that 
it is impossible to build a useful machine without implementing efficient error 
correction. Thus the question in the title is equivalent to asking whether quantum 
computing is possible altogether.

In a previous publication \cite{dyakonov}, I too have accepted the threshold theorem 
but argued that the 
required enormous precision will not be achieved in any foreseeable future. The 
purpose of this article, intended for physicists, is to outline the ideas of quantum 
error correction and to take a look at the technical instruction for fault-tolerant 
quantum computation, first put forward by Shor and elaborated by other mathematicians.
It seems that the mathematics behind the threshold theorem is somewhat detached from 
the physical reality, and that some flawless elements are always present in the 
construction. This raises serious doubts about the possibility of large scale quantum 
computations, even as a matter of principle.

\vskip 1cm

{\bf 2. Brief outline of ideas}\\

The idea of quantum computing is to store information in the values of $2^N$ 
amplitudes describing the wavefunction of $N$ two-level systems, called {\it qubits}, and to 
process this information by applying unitary transformations ({\it quantum gates}), 
that change these amplitudes in a very precise and controlled manner, see the clear 
and interesting review by Steane \cite{steane1}. The value of $N$ needed to have a 
useful machine is estimated as $10^3 - 10^6$. Thus, even for $N=1000$, the number of 
continuous variables (the complex amplitudes of this grand wavefunction) that we are 
supposed to control, is $2^{1000} \sim 10^{300}$. For comparison, the total number of 
protons in the visible Universe is only about $10^{80}$ (give or take a couple of orders of 
magnitude).

The interest in quantum computing surged after Shor \cite{shor1} invented his famous 
algorithm for factorizing very large numbers and showed that an ideal quantum 
computer can solve this problem much faster than a classical computer, which requires 
exponentially great time and resources.

Generally, there are many interesting and useful things one could accomplish with {\it 
ideal} machinery, not necessarily quantum. For example, one could become younger by 
{\it exactly} reversing all the velocities of atoms in our body (and in some near 
environment), or write down the full text of all the books in the world in the  
{\it exact} position of a single particle, or store information in the $10^{23}$ 
vibrational amplitudes of a cubic centimeter of a solid. Unfortunately, unwanted 
noise, fluctuations, and inaccuracies of our manipulations impose severe limits to 
such ambitions. Thus, while the ideas of quantum computing are fascinating and 
stimulating, the possibility of actually building a quantum computer, even in some 
distant future, was met from the start with a healthy scepticism
\cite{unruh,land,har}.

Unlike the digital computer employing basically the on/off switch, which is stable 
against small-amplitude noise, the quantum computer is an analogous machine where 
small errors in the values of the continuous amplitudes are bound to accumulate 
exponentially. So, it seems that a quantum computer of a complexity sufficient to be 
of any practical interest will never work.

In response to this challenge, Shor \cite{shor2} and Steane \cite{steane2} proposed the 
idea of quantum error correction - an ingenious method designed primarily for 
bypassing the so-called "no cloning" theorem: an unknown quantum state cannot be copied.
 At first glance, this theorem prevents us from checking whether there are errors in the 
quantum amplitudes, so that one can correct them. The idea of quantum error correction
 is to spread the information contained in a {\it logical} qubit among several 
 {\it physical} qubits and to apply a special operator, which detects errors in 
 physical qubits (the {\it error syndrome}) and writes down the result by changing 
 the state of some auxiliary {\it ancilla} qubits. By measuring the ancilla qubits 
 only, we can see  the error in the original quantum state and then correct it (see 
 Section 6).
  
However, this method assumes that the ancilla qubits, the measurements, and the 
unitary transformations to be applied, remain ideal. It is said that this type of 
error correction is not fault-tolerant, whatever this may mean. (If the ancilla 
qubits are  flawless, why not use them in the first place?) The ultimate solution, 
the {\it fault-tolerant quantum computation}, was advanced by Shor \cite{shor3} and further 
developed by other mathematicians, see Refs. 10 - 14 and references therein.  Now, 
nothing is ideal: all the qubits are subject to noise, measurements may contain 
errors, and our quantum gates are not perfect. Nevertheless, the threshold theorem 
says that arbitrarily long quantum computations are possible, so long as the errors 
are not correlated in space and time and the noise level remains below a certain 
threshold. In particular, with error correction a single qubit may be stored in memory, 
i.e. it can be maintained arbitrarily close to its initial state during an 
indefinitely long time.

This striking statement implies among other things that, once the spin resonance is 
narrow enough, it can be made {\it arbitrarily} narrow by active intervention with 
imperfect instruments. This contradicts all of our experience in physics.  Imagine a 
pointer, which can rotate in a plane around a fixed axis. Fluctuating external fields 
cause random rotations of the pointer, so that after a certain relaxation time the 
initial position gets completely forgotten. How is it possible that by using only 
other identical pointers (also subject to random rotations) and some external fields 
(which cannot be controlled perfectly), it might be possible to maintain indefinitely 
a given pointer close to its initial position? The answer we get from experts in the 
field, is that it may work because of quantum mechanics: "{\it We fight entanglement with 
entanglement}"\cite{presk} or, in the words of the Quantum Error Correction Sonnet 
by Gottesman \cite{got1},\\

{\it With group and eigenstate, we've learned to fix}
\newline\indent
{\it Your quantum errors with our quantum tricks}.\\

This does look suspicious, because in the physics that we know, quantum-mechanical 
effects are more easily killed by noise than classical ones. Before going into the 
details of the proposed fault-tolerant computation, we present in the following 
section a slight divertissement relevant to our subject.

\vskip 1cm

{\bf 3. Capturing a lion in a desert}\\

The scientific folklore knows an anecdote about specialists in various fields proposing 
their respective methods of capturing a lion in a desert. (For example, the 
Philosopher says that a captured lion should be defined as one who is separated from 
us by steel bars. So, let's go into a cage, and the lion will be captured). Here, we 
are concerned with the Mathematician's method \cite{kostya}:\\

  {\it The desert D being a separable topological space, it contains a countable 
  subset S that is everywhere dense therein. (For example, the set of points with 
  rational coordinates is eligible as  S.)  Therefore, letting $x \in D$ be the point at 
  which the Lion is located, we can find a sequence ${x_n} \subset S$ with $\lim _{n 
  \rightarrow \infty} x_n = x$.  This done, we approach the point $x$ along the sequence ${x_n}$ 
  and capture the lion.} \\

\begin{figure}

\epsfxsize=230pt {\epsffile{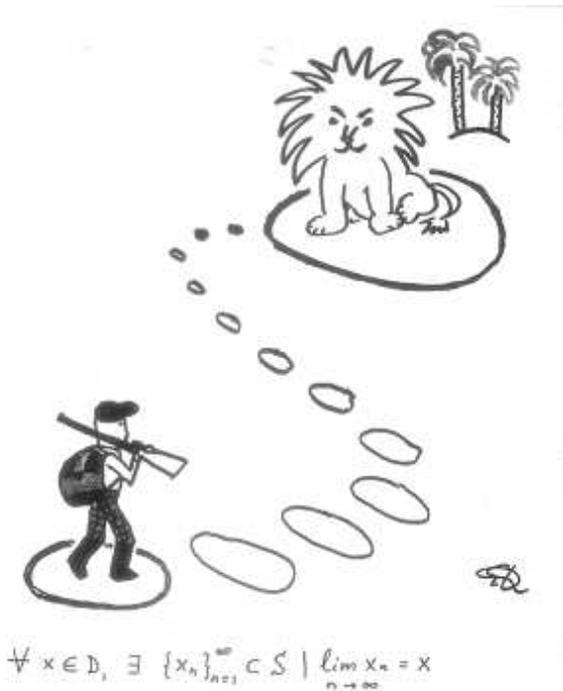}}
\caption{The Mathematician's method of capturing the lion in the desert}

\end{figure}

This method, illustrated in Fig. 1, assumes that the only relevant property of the 
Lion is to be located at a given point in 2D space. Note also that neither {\it time}, 
nor 
what can happen to the Lion and the Hunter during the process, is a point of concern. 
And finally, it is not specified how the sequence ${x_n}$ should be chosen, nor what 
 the limit $n \rightarrow \infty$ could mean in practice. These points are left to be 
 elaborated by the practical workers in the field.

Certainly, mathematics is a wonderful thing, both in itself and as a 
powerful tool in science and engineering.  However, we must be very careful and 
reluctant in accepting theorems, and especially technical instructions, provided by 
mathematicians in domains outside pure mathematics \cite{comment1}. Whenever there 
is a complicated issue, whether in many-particle physics, climatology, or economics, 
one can be almost certain that no theorem will be applicable and/or relevant, because 
the explicit or implicit assumptions, on which it is based, will never hold in 
reality. The Hunter must first explain to the Mathematician what a lion looks like.
\vskip 0.8cm

{\bf 4. Spin relaxation, or decoherence}\\

While the relaxation of two-level systems was thoroughly studied during a large part 
of the 20th century, and is quite well understood, in the quantum computing literature 
there is a strong tendency of mystifying the  relaxation  process and make it look as 
an obscure quantum phenomenon \cite{shor3, presk}: {\it "The qubit (spin) gets 
entangled with the environment..."} or {\it "The environment is constantly trying to 
look at the state of a qubit, a process called  decoherence", etc}. 

In a way, this sophisticated description may be  true, however it is normally quite 
sufficient to understand spin relaxation as a result of the action of a
 time-dependent Hamiltonian $H(t) = {\bf A}(t)\text{\boldmath$\sigma$}$, where 
${\bf A}(t)$ is a random vector function, and $\text{\boldmath$\sigma$}$ are Pauli 
matrices. In simple words, the spin 
continuously performs precession around fluctuating in time magnetic fields. In most 
cases these are not real, but rather effective magnetic fields induced by some 
interactions. A randomly fluctuating field is characterized by its correlation time, 
$\tau _c$, and by the average angle of spin precession, $\alpha$, during time 
$\tau _c$. For the most frequent case when $\alpha << 1$, the spin vector experiences 
a slow angular diffusion. The RMS angle after time $t>>\tau _c$ is $\epsilon \sim 
\alpha(t/\tau _c)^{1/2}$. Hence the relaxation time is $\tau \sim \tau _c /\alpha ^2$.
 If one chooses a time step $t_0$, such that $\tau _c <<  t_0  << \tau _c /\alpha ^2$,
 it can be safely assumed that $\epsilon <<1$, and that rotations during successive 
 time steps are not correlated. 

These random rotations persist {\it continuously} for all the qubits inside the 
quantum computer. It is important to understand that the wavefunction describing an 
arbitrary state of $N$ qubits will deteriorate much faster than each individual qubit 
does. The reason is that this wavefunction
$$ \psi  = A_0 |000...00\rangle + A_1 |000...01\rangle +...+ A_{2^N -1}|111...11\rangle$$
describes complicated correlations between the $N$ qubits, and correlations always 
decay more rapidly. For simplicity, suppose that all qubits are subject to random and 
uncorrelated rotations around the $z$ axis only. Then the state of the $j$th qubit 
will change during one step accordingly to the rule: 
$|0\rangle\rightarrow |0\rangle, |1\rangle \rightarrow
\exp(i\phi _j)|1\rangle$, where $\phi_j$ is the random rotation angle for this qubit. Then 
the amplitudes $A$ will acquire phases $\Sigma \phi _j$, where the sum goes over all 
qubits that are in the state $|1\rangle$ in a given term of $\psi$. The typical RMS value 
of this phase after one step is $\sim \epsilon N^{1/2}$, thus the time it takes for 
unwanted phase differences  between the amplitudes $A$ to become large, is $\sim N$ times 
shorter than the relaxation time for an individual qubit.
 
For this reason, it seems that if we choose the time step so that 
$\epsilon ^2 = 10^{-6}$ (the most cautious of existing estimates for the noise 
threshold), but the quantum state contains $10^6$ qubits, then the computer is likely 
to crash during one step. I am unaware of anybody discussing this problem.

\vskip 0.8cm

{\bf 5. Quantum computation with decoherence-free subspaces}\\

This is a flourishing and respectable branch of quantum computer mathematics (Google 
gives 24 800 hits for "decoherence-free subspaces").  The idea is that there may be 
some special symmetry of the relaxation processes, due to which certain many-qubit 
states do not relax at all. (The simplest model is to consider a relaxation process, 
in which all the qubits are rotated collectively). It is discussed how information 
can be hidden in this decoherence-free subspace, and what would be the best way to 
proceed with quantum computation. The conditions for existence of such subspaces are 
given \cite{lidar} by the following\\

{\bf Theorem 4.} {\it If no special assumptions are made on the coefficient matrix
$a_{\alpha \beta}$ and on the initial conditions $\rho _{ij}$, then a necessary and
sufficient condition for a subspace $\widetilde{\mathcal H}=
\text{\rm Span}[\{|\widetilde k\rangle\}_{k=1}^N]$ to be 
decoherence-free is that all basis states $|\widetilde k\rangle $ are 
degenerate eigenstates of all Lindblad operators} $\{ \text{\boldmath$F$}_\alpha\}$
$$F_\alpha|\widetilde k\rangle=c_\alpha|\widetilde k\rangle\quad
\forall\alpha,k.$$ 
This gives the reader an idea of what the quantum computing literature looks like.

Although it is not difficult to construct artificial models with some special 
symmetries, my guess is that in any real situation the Lindblad operators do not have 
common eigenstates at all.  Obviously, the simplest way to fight noise is to suppose 
that at least {\it something is ideal} (noiseless). Unfortunately, this is not what 
happens in the real world.

\vskip 1cm

{\bf 6.    Quantum error correction by encoding}\\
   
Below is a simplified example \cite{steane3,polak} of the quantum error correction 
using encoding. The simplification results from the assumption that the only errors 
allowed are rotations around the $x$ axis, described by the matrix $E=\cos(\theta /2)I 
- i\sin(\theta/2)\sigma_x$, where $I$ is the unit matrix and $\sigma _x$ is the Pauli 
matrix. For small rotation angles $\epsilon =2\theta << 1$, this gives $E = I - 
i\epsilon \sigma _x$.  In the case when these are the only errors for individual 
qubits, it is sufficient to encode the logical $|0\rangle$ and $|1\rangle$  by three 
physical qubits: $|0\rangle \rightarrow |000\rangle, |1\rangle \rightarrow 
|111\rangle$.  The error correction procedure requires the following steps:\\

1) The general state of a qubit,  $a|0\rangle + b|1\rangle$, is encoded as 
$\psi = a|000\rangle + b|111\rangle$. 
Suppose that the three physical qubits experience small and uncorrelated rotations 
$E_1, E_2$, and $E_3$. Let us see how the initial state can be recovered. For example, 
suppose there is an error in the second qubit only. The wave-function becomes:
$$E_2\psi = [a|000\rangle + b|111\rangle] - i\epsilon _2[a|010\rangle + b|101\rangle].$$

      2) We now mechanically add 3 auxiliary ancilla qubits in the state $|0\rangle$, 
      obtaining the  state:  
      $$E_2\psi = [a|000\rangle + b|111\rangle]|000\rangle - 
 i\epsilon _2[a|010\rangle + b|101\rangle]|000\rangle .$$

      3) We next introduce the syndrome extraction  operator, $S$, defined as:\\

$S|000\rangle|000\rangle =|000\rangle|000\rangle,$ 
\newline\indent   
$S|111\rangle|000\rangle = |111\rangle|000\rangle,$
\newline\indent
$S|100\rangle|000\rangle =|100\rangle|100\rangle,$ 
\newline\indent  
$S|011\rangle|000\rangle = |011\rangle|100\rangle,$
\newline\indent
$S|010\rangle|000\rangle =|010\rangle|101\rangle,$ 
\newline\indent 
$S|101\rangle|000\rangle = |101\rangle|101\rangle,$ 
\newline\indent
$S|001\rangle|000\rangle =|001\rangle|001\rangle,$ 
\newline\indent  
$S|110\rangle|000\rangle = |110\rangle|001\rangle.$\\

The first 3 qubits, containing the data, are left intact. If one of them is 
flipped, then the ancilla bits are changed accordingly. The operator $S$ writes 
down the error into the ancilla, allowing us to identify the error location. 
Now:  
$$SE_2\psi = [a|000\rangle + b|111\rangle]|000\rangle -
 i\epsilon _2[a|010\rangle + b|101\rangle]|101\rangle.$$

4) Finally, we measure the three ancilla qubits.  If we get $(000)$, then we do 
nothing, since  this result shows that the state automatically has been reduced to 
the initial state $\psi = a|000\rangle + b|111\rangle$. If (with a small probability equal to 
$\epsilon_{2} ^ 2$) we obtain the result $(101)$, then we understand that there is an 
error in the 2nd qubit, and that the state of the remaining (data) qubits is  
$a|010\rangle + b|101\rangle$. This error is corrected by applying the operator 
$\sigma_x$  to the second qubit.  Thus, in both cases we recover the original state 
$$\psi = a|000\rangle + b|111\rangle \rightarrow a|0\rangle + b|1\rangle.$$
The method works equally well if all three data qubits have errors, provided that 
second order terms in the $E_1 E_2 E_3 \psi$ wavefunction proportional to
$\epsilon_1 \epsilon_2$, $\epsilon_1 \epsilon_3$, and $\epsilon_2 \epsilon_3$  are 
neglected.  This is justified by the small probability for the admixture of the 
corresponding states, proportional to $\epsilon^4$ (compared to the $\epsilon^2$  
probability of one-qubit errors). However, this method does not work if errors in 
different qubits are correlated, {\it i.e.} if there is an admixture of states with 
two errors with an amplitude $\epsilon$ (not $\epsilon^2$).  This is why the 
requirement that errors are uncorrelated is crucial.

There is, indeed, a remarkable "quantum trick" here: the (ideal) measurement of 
the ancilla qubits automatically reduces the wavefunction representing a large 
superposition of states to only one of its terms! Because of this, knowing how to 
correct a bit-flip $(|0\rangle \rightarrow |1\rangle$, $|1\rangle \rightarrow |0\rangle)$, referred to as 
"fast error", we can also correct "slow errors", $E = I - i\epsilon \sigma_x$ , for 
arbitrary (but small) values of $\epsilon$. This property is called 
"digitization of noise".

To correct general one-qubit errors a more sophisticated encoding
\cite{shor2,steane2} by a greater 
numbers of qubits is needed. For example, the logical $|0\rangle$ might be encoded as:\\
 
$|0\rangle \rightarrow (1/\sqrt{8} [|0000000\rangle + |0001111\rangle + 
|0110011\rangle + |0111100> + |1010101\rangle + |1011010\rangle + |1100110\rangle + 
|1101001\rangle].$\\

However the principle of error 
correction is the same. It is believed that it may be advantageous to use 
{\it concatenated} encoding, in which each encoding qubit should be further encoded 
in the same manner, and so on... It is supposed that the future quantum engineer
\cite{comment2}  might wish to encode the logical $|0\rangle$ and   $|1\rangle$ by 
complicated superpositions of the states of $7^3 =343$ physical qubits!

What if we apply the same method of error correction not just to one qubit, but to 
some $N$-qubit state? Making exactly the same assumptions, we encode each of the $N$ 
logical qubits by three physical ones, add an appropriate number of ancilla qubits, 
and proceed in the same way as above. Then, in accordance with the remark at the end 
of Section 4, the condition for the method to work will be $N\epsilon^2 << 1$, not 
simply $\epsilon^2 << 1$. Indeed, after applying the syndrome extraction operator, 
the number of one-qubit error terms is $N$ with probabilities $\epsilon^2$, while 
the number of two-qubit error terms is $N(N-1)/2$ with probabilities $\epsilon^4$.  
In order that  the method can work, the total probability of obtaining any two-error 
state during measurements should be small, which is true when $N\epsilon^2 << 1$.

\vskip 1cm

{\bf 7.    The imperfect two-qubit gate}\\

For quantum computation one needs to apply one-qubit gates, but also two-qubit and 
three-qubit gates.  Application of two-qubit gates is necessary from the outset to 
perform the first step, encoding. The problem is that not only individual qubits are 
subject to relaxation, as described in Section 4, but also all of the quantum gates 
are not perfect, because neither the Hamiltonian that should be switched on at the 
desired moment, nor the duration of its action can be controlled exactly. While an 
error in the one-qubit gates can be simply added to the random rotations that exist 
anyway, errors in the two-qubits gates require more care.

We should first decide what an imperfect two-qubit gate is. It seems  that the 
generally accepted model is the following \cite{shor3} :  {\it "For the error model 
in our quantum gates, we assume that with some probability $p$, the gate produces 
unreliable output, and with probability $1-p$, the gate works perfectly."} 

A more detailed description \cite{steane1,steane3} specifies what exactly an 
unreliable output is: {\it "The failure of a two-qubit gate is modeled as a process 
where with probability   $1-\gamma_2$  no change takes place before the gate, and 
with equal probabilities $\gamma_2 /15$  one of the 15 possible single- or two-qubit 
failures take place." } 

In other words, the faulty gate is supposed either to act as an ideal one (with high 
probability), or as an ideal gate preceded by uncorrelated errors in the two qubits 
involved (with low probability). In reality, there will always be some more or less 
narrow probability distribution around their desired values of the 16 real 
parameters defining the unitary transformation. Never, under any circumstances, 
will an ideal gate exist. The crucial difference is that any real gate will 
introduce correlated errors of the two qubits. Such correlated errors are not 
correctable within the error-correcting scheme described in Section 6.

Here is a more sophisticated model \cite{kitaev}: {\it "The noise model we will 
consider can be formulated in terms of a time-dependent Hamiltonian $H$ that governs 
the joint evolution of the system and the bath. We may express $H$ as  $H = H_S + H_B 
+ H_{SB}$, where $H_S$  is the time dependent Hamiltonian of the system that 
realizes the ideal quantum circuit, $H_B$  is the arbitrary Hamiltonian of the bath, 
and $H_{SB}$ couples the system to the bath."} 

This is a quite general approach, especially if the arbitrary (?) Hamiltonian of the 
"bath" describes also the electronic equipment and the quantum engineer himself. 	
However, in reality there is no such thing as a "time dependent Hamiltonian of the 
system that realizes the ideal quantum circuit", like there is no such thing as 
square root of 2 with all the infinite number of its digits. True, such abstractions 
are routinely used in mathematics and theoretical physics. However the whole issue 
at hand is to understand whether the noisy nature of the {\it real} Hamiltonian does, 
or does not, allow to realize anything sufficiently close to the "ideal quantum 
circuit".  Thus, it could well happen that the supposed success of fault-tolerant 
quantum computation schemes is entirely due to the uncontrolled use of 
innocent-looking abstractions and models (see Section 3). A very careful analysis 
is needed to understand the true consequences of any simplifications of this kind.

Another implicit assumption, which may be not quite innocent, is that the gates are 
infinitely fast. In fact, {\it during} error correction new errors may appear 
\cite{alicki}.

\vskip 0.8cm

{\bf 8.    The prescription for fault-tolerant quantum computation}\\

The error correction scheme, briefly described in Section 6, assumes that encoding, 
syndrome extraction, and recovery, are all ideal operations, that ancilla qubits are 
error-free, and that measurements are exact. Fault-tolerant methods are based on the 
same idea, but are supposed to work even if all these unrealistic assumptions 
(making, in fact, error correction unnecessary) are lifted. The full instructions 
\cite{shor3,presk,got1,steane3,steane4} are extraordinary complicated, details 
that may be important are often omitted, and the statements are not always quite 
clear. The basic ideas are as follows.\\
 
1) There exists a universal set of three gates, sufficient for quantum computation. 
{\it "The proof of this involves showing that these gates can be combined to produce 
a set of gates dense in the set of 3-qubit gates"} \cite{shor3} (see also Section 3).  
In other words, 
any gate can be approximated to any desired accuracy by application of a large 
enough number of the three special gates belonging to the universal set.\\

2) These three gates can be used in a fault-tolerant manner, which means in such a 
way that only uncorrelated, and thus correctable, errors are produced. 
Fault-tolerance is achieved by encoding the logical qubits, using specially prepared 
states of ancilla qubits, and some rules designed to avoid error propagation. Thus 
application of a single 3-qubit gate fault-tolerantly amounts to a mini-  quantum 
computation with thousands of elementary operations and intermediate measurements.\\

3) The encoding itself cannot be done fault-tolerantly: {\it "Therefore, 
we should carry out a measurement that checks that the encoding has been done 
correctly. Of course, the verification itself may be erroneous, so we must repeat 
the verification a few times before we have sufficient confidence that the encoding 
was correct"} \cite{presk}.

A similar prescription for verification concerns ancilla states: {\it 
"However, the process of creating the ancilla blocks may introduce correlated errors,
 and if those errors enter the data, it will be a serious problem. Therefore, we 
 must also verify the ancilla blocks to eliminate such correlated errors. Precisely 
 how we do this is not important for the discussion below, but it will certainly 
 involve a number of additional ancilla qubits"} \cite{got2}.
	
It is recommended \cite{shor3,presk} to construct auxiliary "cat states":   
$(|00...00\rangle + |11...11\rangle)/\sqrt{2}$, where the size of the cat depends 
on the number of qubits used for encoding. Again, since there may be (one might 
better say: there {\it always will be}) errors in the cat state, it must be verified 
before being used.\\ 
    
4) Operations should be repeated:  {\it "If we mistakenly accept the measured 
syndrome and act accordingly, we will cause further damage instead of correcting the 
error. Therefore, it is important to be highly confident that the syndrome 
measurement is correct before we perform the recovery. To achieve sufficient 
confidence, we must repeat the syndrome measurement several times"} \cite{presk}.\\
      
5) All these precautions still do not guarantee that the computer will not crash. 
However, what matters is the {\it probability} of crash. Once this probability is 
small enough, which is supposed to happen at a low enough noise rate, we can repeat 
the whole quantum calculation many times to get reliable results. By estimating the 
crash probability, one obtains an estimate for the threshold noise level.\\
      
A detailed description of the fault-tolerant computation rules can be found in 
Ref. 10. I don't find this description clear and/or convincing enough. Taking
properly in account the continuous nature of random qubit rotations, gate 
inaccuracies, and measurement errors, even with all the verifications and 
repetitions, there seems to be no way to avoid small admixtures of unwanted states.
 
In fact, the pure spin-up state can never exist in reality (one reason is that we never 
know the exact direction of the $z$ axis).  Similarly in the classical world we can never 
have a pointer looking exactly in the $z$ direction.  Generally, no desired state can 
ever be achieved exactly, rather, whatever we do, we will always have an admixture of 
unwanted states, more or less rich. One can never have an exact 
$(|0\rangle + |1\rangle)/\sqrt{2}$ state, let alone more complicated "cat" states like 
$(|0000000\rangle + |1111111\rangle)/\sqrt{2}$. Such abstractions must be used with 
extreme caution when discussing the role of errors and inaccuracies.
 
When the small undetected and unknown admixture of unwanted states together with the 
"useful" state is fed into the subsequent stages of the quantum network, it is most 
likely that the error will grow exponentially in time. Accordingly, the crash time 
will depend only logarithmically on the initial error value. This is what happens 
when one tries to reverse the evolution of a gas of hard balls. At a given moment 
one reverses the direction of all the velocities, but oops, the gas  never returns 
to its initial state (even in computer simulation, let alone reality). The reason is 
that however small the initial (and subsequent) errors are, they will increase 
exponentially (the Lyapunov exponent). It is a great illusion to think that things 
are different in quantum mechanics.

Related to this, there is another persistent misunderstanding of quantum mechanics, 
which plagues the quantum error correction literature. Using quite classical 
language, one says that the qubit "decoheres" with probability $p=\sin ^2 \theta$, 
instead of saying: the qubit is in the state $\psi = \cos \theta |0\rangle + 
\sin \theta |1\rangle$. It makes only a semantic difference if we are going to immediately 
{\it measure} the qubit, since the probability of finding it in a state $|1\rangle$ is 
indeed $p$. However, this language becomes completely wrong if we consider some further 
evolution of 
our qubit with a unitary matrix $R$. The common thinking (applied, for example, for 
estimating the noise threshold) is that we will have the state $R|0\rangle$ with 
probability $1-p$, and the state $R|1\rangle$ with probability $p$. In reality, we will 
have the state $R\psi$, and it is not the same thing. The former line of reasoning 
gives the probability of measuring $|0\rangle$ in the final state as 
$(1-p)|\langle0|R|0\rangle|^2 + p|\langle0|R|1\rangle|^2$, while the latter (and correct) 
one will give $|\langle0|R|\psi\rangle|^2$, and 
these results are very different. As an exercise, the reader can take for $R$ a 
rotation of our qubit around the $x$ axis by some angle and compare the results. 
A quantum-mechanical surprise lies in store. In quantum mechanics, one cannot 
calculate probabilities by considering what happens to some ideal states. Instead, 
one must look at the evolution of the real states which always differ from ideal ones 
by some admixture of unwanted states.

Another point is that the finite time needed to do anything at all, is usually not 
taken into account (see Section 3). According to the procedure described in Section 
6, measuring the syndrome and obtaining $(000)$ indicates the correct state that 
requires no further action.  In fact, while we were making our measurements, the 
data qubits have experienced their random rotations. And no matter how many times we 
repeat the syndrome measurements this will happen again and again. So why bother 
with error correction? 

Alicki \cite{alicki} has made a mathematical analysis of the consequences of finite 
gate duration. I am not in a position to check his math, but I like his result: the 
fidelity exponentially decreases in time. He writes: {\it "...unfortunately, the 
success of existing error correction procedures is due to the discrete in time 
modelling of quantum evolution. In physical terms discrete models correspond to 
unphysical infinitely fast gates".}

\vskip 0.8cm

{\bf 9.    Designing perpetual motion machine of the second kind}\\

This is certainly {\it not} equivalent to achieving fault-tolerant quantum 
computation, during which we will put some energy into the system by applying 
external fields and performing measurements.

However there is a certain similarity between the two problems in the sense that 
what we are trying to do is to maintain a reversible evolution of a large system 
with many degrees of freedom in the presence of noise and using noisy devices
\cite{comment3}. People, who have had the opportunity of considering projects of 
perpetual motion machines, know their basic principle: insert at least {\it one} 
ideal ({\it i.e.} not sensitive to thermal fluctuations) element somewhere deep 
within a complicated construction. Finding and identifying such an ideal element may 
be a daunting task. 

Naively, one starts with proposing a valve that preferentially lets through only 
fast molecules.  Next, one understands, that the valve itself is "noisy", so that it 
will not work as expected. However, if one adopts the noise model, in which the 
valve is faulty with probability $p$ but works perfectly with a probability $1-p$, 
or makes a sophisticated construction involving many valves connected by wheels and 
springs, and if just one of these elements is considered as {\it ideal} (or even 
working perfectly with some probability), one can immediately arrive at the 
conclusion that a perpetual motion machine feeding on thermal energy is possible.

This lesson should make us extremely vigilant to the explicit or implicit presence 
of ideal elements within the error-correcting theoretical schemes.

\vskip 0.8cm

{\bf 10.       Challenge}\\

After ten years of doing mathematics devoted to fault-tolerant quantum computation, 
maybe the time is ripe for making a simple numerical test. Let us focus on the 
simplest, almost "trivial" task of storing just one qubit and let us verify the 
statement that its initial state can be maintained indefinitely in the presence of 
low-amplitude noise. Take for example an initial state $(|0\rangle + |1\rangle)/\sqrt{2}$, 
a spin pointing in the $x$ direction.

All qubits experience continuous random uncorrelated rotations, the RMS rotation 
angle being small during one time step. As a further simplification, one could 
restrict rotations to the $xy$ plane only. The two- and three-qubit gates have a 
narrow probability distribution for all of the parameters, defining the 
corresponding unitary transformation, around their desired values. Errors in 
successive gates are not correlated. We have a refrigerator containing an unlimited 
number of ancilla qubits in the state $|0\rangle$. Once the ancillas are out of the 
refrigerator, they become subject to the same random rotations. Measurements involve 
errors: when the state $a|0\rangle + b|1\rangle$ is measured with the result $(0)$, 
the quantum state is {\it not} reduced to exactly $|0\rangle$, but rather to $|0\rangle + 
c|1\rangle$ with some unknown, but hopefully small $c$. Allocate a certain time for 
measurements and duration of gates and take into account that all qubits continue to 
be randomly rotated during this time. This simplest model cannot be relaxed further 
without entering an imaginary world where something is ideal.

Presumably, to maintain our single qubit close to its initial state, a certain 
sequence of operations (with possible branching depending on the result of 
intermediate measurements) should be applied periodically. {\it Provide a full list of 
these elementary operations}, so that anybody can use a PC to check whether qubit 
storage really works. The future quantum engineer will certainly need such a list! 
{\it If} it works, this demonstration would be a convincing, though partial, proof 
that the idea of fault-tolerant quantum computation is sound.
 
Steane \cite{steane3} undertook a thorough numerical simulation of error propagation 
in a quantum computer with results confirming the threshold theorem. Since an exact 
simulation of quantum computing on a classical computer is impossible, he used a 
partly phenomenological model based on the (questionable) assumption that {\it "it is 
sufficient to keep track of the propagation of errors, rather than the evolution of 
the complete computer state"}. Since the real errors always consist in admixtures of 
unwanted states, considering the quantum evolution of the complete computer state 
seems to be the only way to respect quantum mechanics (see  Section 8).

The task of maintaining a single qubit in memory is  much simpler and, once the list 
of required operations is provided, it hopefully can be simulated in a 
straightforward manner. I predict that the result will be negative.

\vskip 0.8cm

{\bf 11.     Conclusion}\\

It is premature to accept the threshold theorem as a proven result. The state of a 
quantum computer is described by the monstrous wavefunction with its $10^{300}$ 
complex amplitudes, all of which are continuously changing variables. If left alone, 
this wavefunction will completely deteriorate during $1/N$ of the relaxation time of 
an individual qubit, where $N \sim 10^3 - 10^6$  is the number of qubits within the 
computer. 

It is absolutely incredible, that by applying external fields, which cannot be 
calibrated perfectly, doing imperfect measurements, and using converging sequences 
of "fault-tolerant", but imperfect, gates from the universal set, one can 
continuously repair this wavefunction, protecting it from the random drift of its 
$10^{300}$ amplitudes, and moreover make these amplitudes change in a precise and 
regular manner needed for quantum computations. And all of this on a time scale 
greatly exceeding the typical relaxation time of a single qubit.

The existing prescriptions for fault-tolerant computation are rather vague, and the 
exact underlying assumptions are not always clear. There are several subtle issues, 
some of which were discussed above, that should be examined more closely. It seems 
likely that the (theoretical) success of fault-tolerant computation is due not so 
much to the "quantum tricks", but rather to the backdoor introduction of ideal 
(flawless) elements in an extremely complicated construction. Previously, this view 
was expressed by Kak \cite{kak}.

It would be useful to check whether the fault-tolerant methods really work by 
numerically simulating the quantum evolution during the proposed recovery procedures 
for a single qubit using a  realistic noise model, which does not contain any ideal 
elements. \\


\begin{thebibliography}{}
\bibitem{comment1} Somebody has proved a theorem concerning the existence of bound 
states for the one-dimensional Schroedinger equation. The surprise comes for  
repulsive potentials $U(x) = - Ax^n$ : it is proved that for $n=4,6...$ bound states 
always exist. The reader might be interested to find out a)why this theorem is true 
b)what is its physical meaning and c)why it is  irrelevant for the real world.
\bibitem{comment3} Because of this similarity, it is quite probable that the design 
and the theory of such machines will become the next hot topic, especially if a good 
name (beginning with the magic word "quantum") for this activity will be invented.
\end{thebibliography}
\end{document}